\documentclass[conference]{ieeeconf}
\usepackage[english]{babel}
\IEEEoverridecommandlockouts
\let\labelindent\relax
\usepackage{times}
\usepackage{geometry}
\usepackage[utf8]{inputenc}
\usepackage{enumitem,amssymb}
\usepackage{ragged2e}
\newlist{thematic}{itemize}{8}
\setlist[thematic]{label=$\square$}
\usepackage{pifont}
\usepackage{titlesec}
\usepackage{amsmath}
\titleformat{\section}{\large\bfseries}{\thesection}{1em}{}

\usepackage{setspace}
\usepackage{comment}
\usepackage[utf8]{inputenc}

\usepackage{color}
\usepackage{fancyhdr}
\usepackage{lastpage}
\usepackage{multirow}

\usepackage{verbatimbox}
\usepackage{subcaption}
\usepackage{graphicx}
\usepackage{multicol}
\usepackage{graphicx}
\usepackage{epstopdf}
\usepackage{floatrow}
\usepackage{subfig}
\usepackage{gensymb}
\usepackage{listings}
\usepackage{mcode}

\usepackage{xparse}

\usepackage[font={small,sf},labelfont=bf]{caption}

\IEEEoverridecommandlockouts                              

\begin{document}

\title{\LARGE \bf
Embodied Supervision: Haptic Display of Automation Command to Improve Supervisory Performance
}

\author{Alia Gilbert$^{1}$, Sachit Krishnan$^{1}$, and R. Brent Gillespie$^{1,2}$
\thanks{$^{1}$Department of Robotics, University of Michigan,}
\thanks{$^{2}$Department of Mechanical Engineering, University of Michigan.}
}

\maketitle
\thispagestyle{empty}
\pagestyle{empty}

\begin{abstract}

A human operator using a manual control interface has ready access to their own command signal, both by efference copy and proprioception. In contrast, a human supervisor typically relies on visual information alone. We propose supplying a supervisor with a copy of the operator’s command signal, hypothesizing improved performance, especially when that copy is provided through haptic display. We experimentally compared haptic with visual access to the command signal, quantifying the performance of N=10 participants attempting to determine which of three reference signals was being tracked by an operator. Results indicate an improved accuracy in identifying the tracked target when haptic display was available relative to visual display alone.  We conjecture the benefit follows from the relationship of haptics to the supervisor's own experience, perhaps muscle memory, as an operator.

\end{abstract}

\section{Introduction}

As the capabilities of automation advance, humans are promoted from the role of operator to supervisor, often being asked to monitor multiple automated agents simultaneously. As supervisors, humans are expected to detect automation faults, to intervene when recovery is beyond automation capabilities, and to re-program automation objectives when necessary. Yet humans are notoriously ill-equipped to supervise \cite{bainbridge1983ironies}. Humans lose vigilance when sustained attention is required \cite{sheridan2012human} and have difficulty interpreting the visual displays conventionally used in supervisory control settings \cite{chen2010supervisory} \cite{sheridan2016human}.

When considering differences in the roles of operator and supervisor, an opportunity to provide the supervisor with additional information becomes apparent. An operator generates and therefore has access to the command signal $u(t)$ that they issue through the manual control interface, by virtue of both efference copy \cite{bridgeman2007efference}, \cite{imamizu2010prediction} and proprioception. A supervisor, on the other hand, relies strictly on visual monitoring of the reference signal $r(t)$ and system output $y(t)$ \cite{sheridan2012human}.  The question arises, would the control signal $u(t)$ be of any benefit to the supervisor? 

The privilege afforded the operator (the person in the driver's seat, whether literally or figuratively) is well known. For example, a driver rarely suffers motion sickness like the vehicle passengers, precisely because the driver can anticipate visual and/or vestibular cues \cite{rolnick1991driver}. 
In human motor behavior, the advantage of being in control has long been recognized \cite{von1971principle}. 
By holding an \emph{internal model} of the system under control, and having generated the control command $u(t)$, the operator can produce an \emph{expected} sensory feedback signal. Von Holst and Mittelstaedt postulated that only when discrepancies arise between expected and actual sensory feedback do higher level control systems need to be involved. Expected sensory feedback can be compared to the sensory input that results from human movement (called reafference) \cite{von1971principle}, enabling a shielding of perception from self-induced effects on the sensory stream. That shielding is critical for stabilizing the sensory field, for example in removing the effects of eye and head movements from the image projected on the retina \cite{jeannerod2003action}.

It seems plausible then, that if the supervisor has a copy of $u(t)$, the same benefits afforded the operator might also accrue for the supervisor. 
By placing a manual control interface that moves under the action of another agent into the passive hand of a human supervisor (who does not generate the control signal), we hypothesize that the supervisor's ability to anticipate the response $y(t)$ will improve.  We also hypothesize that the supervisor will be in a better position to determine the \emph{control intent} of the operator.

Determining the control intent of an operator is particularly important when one shares control with another operator. 
Haptic Shared Control \cite{griffiths2005sharing,abbink2012haptic} is in part an example of displaying an autonomous agent's command $u(t)$ for the purpose of giving access to a human driver. By feel, the driver senses the automation's command torque.
 However, Haptic Shared Control is fundamentally a paradigm for \emph{control sharing } whereas the paradigm we address in this paper is \emph{supervisory control}, in which the transmission of information is strictly from autonomous agent to human supervisor. 
 The supervisor is not able to ``edit'' the automation's command $u(t)$ as in Haptic Shared Control---though the supervisor may choose to \emph{take over} control from the automation (in which case the supervisory role is suspended).


We conjecture that prior experience acting as an operator can come into to play in the supervisor's perceptual processing of the command signal $u(t)$ generated by another agent.  Further, we expect that such sensorimotor processing will be more aligned with a haptic than a visual copy of $u(t)$.  Haptics may carry advantage for supplying the command copy
in a form recognizable to the supervisor as a signal they might themselves have issued.

In this paper we describe an experiment designed to quantify the merits of providing a copy of an operator's command $u(t)$ to a supervisor/participant.  We devised an experiment in which the command of a human operator (``wizarding'' an automation system) was either not shared with the experiment participant, shared visually, or shared haptically. The participant acted as a supervisor in all cases.  Before presenting the experimental methods and results, we describe a simple simulation study that interprets the supervisor/participant as an input observer. The paper closes with a brief discussion of results and future work.



\section{Preliminaries}
\begin{figure}
    \centering
    \includegraphics[width=1\linewidth]{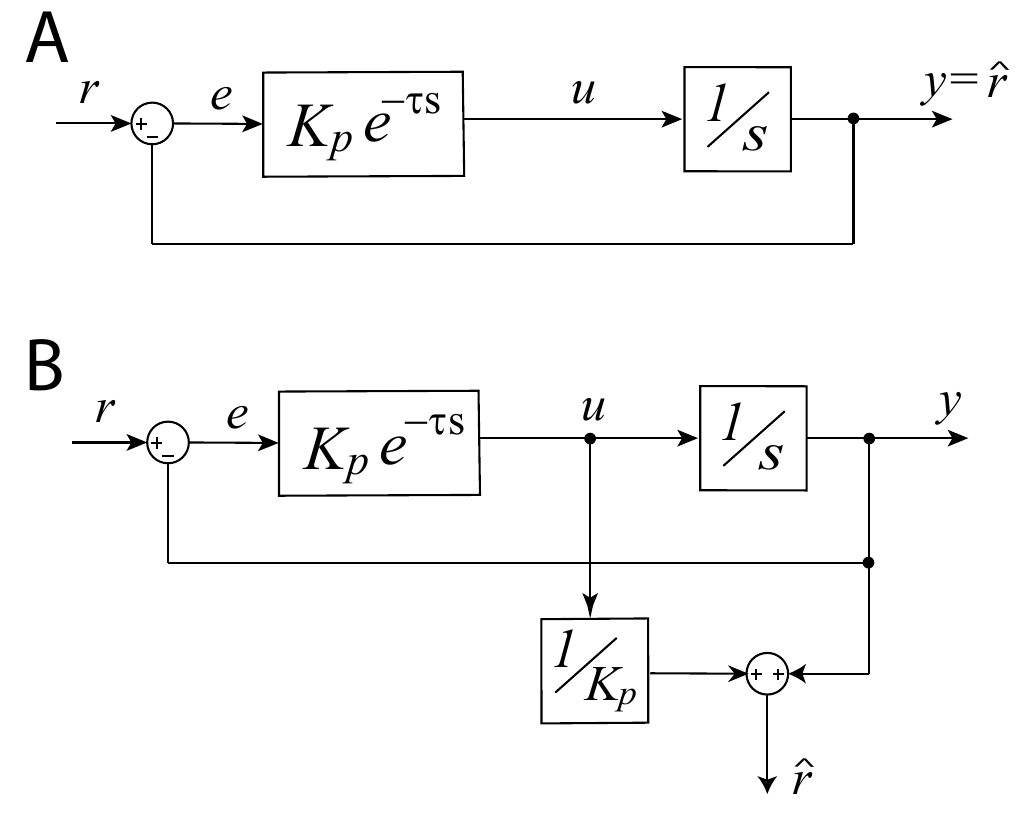}
    \caption{\textbf{Block Diagram:} (A): The supervisor without access to $u(t)$ must use $y(t)$ as the best estimate of $r(t)$, whereas (B): a supervisor able to observe $u(t)$ can construct a superior estimate $\hat{r}(t)$ by combining a scaled copy of $u(t)$ with $y(t)$.  }
    \label{fig:block_diagram}
\end{figure}


The McRuer Crossover model is perhaps the most common example of a human control model for tracking tasks \cite{mcruer1965human}. This validated model predicts that humans adapt their control behavior to the dynamics of a controlled element to produce an integrator in the open-loop transfer function when tracking random-appearing reference trajectories. For example, in the case of a controlled element  $Y_c = \frac{K_c}{s}$, the corresponding human operator transfer function, according to McRuer, is $Y_p(j\omega) = K_pe^{-j\omega \tau_e}$. We use this model in a simple simulation study to show how displaying the command signal to a human observer can improve their estimation of the reference signal $r(t)$. 

 We assess supervisor performance as the supervisor watches the system output $y(t)$ among a field of possible targets $r(t)$ being tracked. That is, the supervisor is tasked with determining which of several targets is being tracked by an operator controlling the system output. Typically, a supervisor does not have access to the command signal $u(t)$ produced by the operator and imposed on the plant. In effect, we seek the simplest competent model of the human as a supervisor under two conditions: one with and the other without access to $u(t)$. 

A block diagram showing a feedback controller $C(s) = K_p e^{-\tau s}$ acting on an integrator plant $P(s) = \frac{1}{s}$ is presented in Fig. \ref{fig:block_diagram}. This controller models a human operator performing a compensatory tracking task, according to the McRuer crossover model. We set the crossover frequency $\omega_c = 4.3$ rad/s and delay $\tau = 0.14 s$ using the values tabulated in \cite{mcruer1967review,mcruer1974mathematical}. The reference signal $r(t)$ appears as a sum of sinusoids. While in applications we presume that a supervisor is tasked with supervising an automation system rather than a human operator, we chose the McRuer Crossover model as a plausible model of an automation system. Indeed, in the experiment described below, we use a human operator to ``wizard'' an automation system. 

The upper block diagram in Fig. \ref{fig:block_diagram} depicts the case in which the human supervisor lacks access to $u(t)$ and must rely on the signal $y(t)$ and the assumption that the operator is performing well tracking $r(t)$ to come up with a best estimate $\hat{r}(t)$ of the signal being tracked. Thus the supervisor uses $y(t)$ itself as the best estimate of $r(t)$. 
The lower block diagram in Fig. \ref{fig:block_diagram} depicts a supervisor who has access to the signal $u(t)$ in addition to $y(t)$. In this case the supervisor may use the simple construction $\hat{r}(t) = \frac{u(t)}{K_p} + y(t)$ to come up with an estimate of the signal $r(t)$ being tracked. Such an estimate will be superior to $y(t)$. 


  
\begin{figure}[h!]
    
    \centering
    \includegraphics[width=1\textwidth]{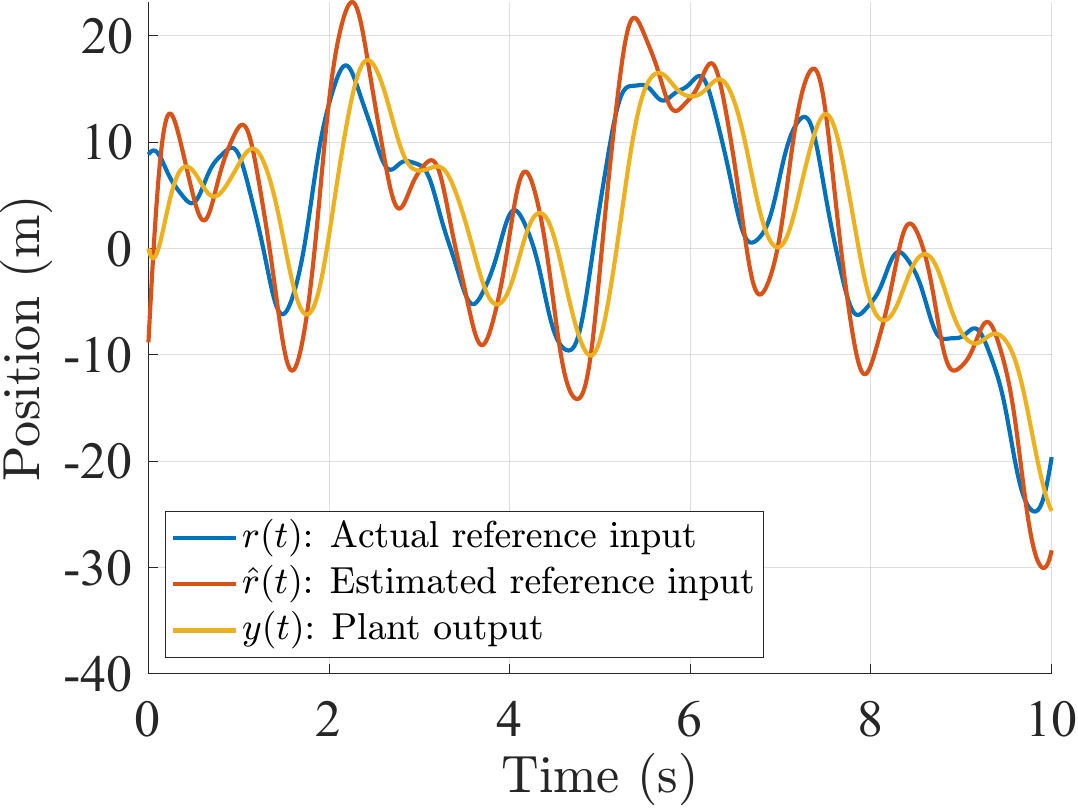}
    \label{fig:sim_results_a}
\caption{\textbf{Simulation Results:} When the supervisor does not have access to $u(t)$, their best estimate of the reference signal $r(t)$ (blue) is $y(t)$ itself (yellow). On the other hand, when the supervisor has access to $u(t)$, their best estimate of $r(t)$ is $\hat{r}(t)$ (red). As is evident in the plots, $\hat{r}(t)$ has a much smaller phase difference with $r(t)$ than $y(t)$ does, suggesting that $\hat{r}(t)$ is more effective at accounting for the controller's delay and phase lag in the integrator plant.}  
\label{fig:sim_results}
\end{figure}

 Qualitatively speaking, without access to the control command, the supervisor uses $y(t)$ as their estimate of $r(t)$ and therefore has to rely on the operator to track the target closely to be able to identify to the correct target. However, when the supervisor does have access to the control command, they have access to a signal that is a $90\degree$ phase ahead of the output. 
  This places the supervisor in a position to make better (and earlier) estimates of which reference signal the autonomous agent is tracking. These phase differences are clearly visible in Fig \ref{fig:sim_results}. 
  The average Root Mean Square (RMS) errors of $y(t)$ and $\hat{r}(t)$ with respect to $r(t)$ are $0.6645$ m/s and $0.4629$ m/s, respectively, indicating that $\hat{r}(t)$ is a better estimate of $r(t)$ than $y(t)$. These models are extensible to more modern Kalman Filter-based descriptions for human motor behavior, which is our plan for immediate future work.  We expect the simplest competent model highlighted here can be used to generate additional hypotheses about human supervision of autonomous agents controlling plants with various dynamical features (pure delay, rate control, acceleration control, typical vehicle dynamics, and so on).

\section{Methods}

\begin{figure}[h]
\caption{Experimental Setup: The Operator Station is on the left and Supervisor/Participant Station is on the right. }
\centering
\includegraphics[width=1\textwidth]{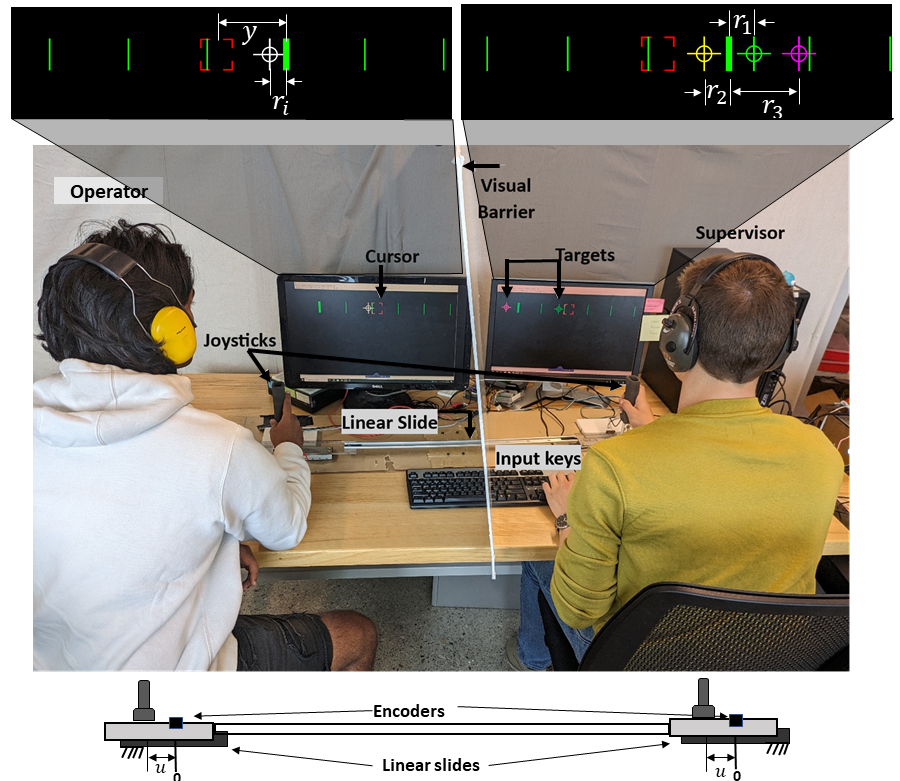}
\label{fig:experimental_setup}
\end{figure}

\subsection{Participants}
Ten persons participated in the experiment, of which
four were women, each were between twenty and twenty-six
years old, and all reported right hand dominance. None of the participants reported trauma on the tested hand that could have affected his/her motor function or haptic perception. 

\subsection{Experimental Setup}

 We used a simple apparatus consisting of two identical handles (joysticks) mounted independently on low friction linear slides (see Fig. \ref{fig:experimental_setup}). Each linear slide was instrumented with a linear encoder (US Digital, 250LPI).  A computer monitor was placed behind each joystick and a visual barrier was placed so that a human user sitting in front of given joystick could only see the associated monitor.  The left joystick and monitor functioned as the \emph{operator station} and the right joystick and monitor functioned as the \emph{supervisor/participant station}. 
The two joysticks were optionally connected through a bar linkage. When the bar was connected, the two joysticks moved together as a rigid body. When the bar was disconnected, each joystick moved independently.


The joystick at the operator's station was used to control the horizontal location of a red \emph{cursor} shown on both the operator's and the supervisor/participant's monitors. Between the operator joystick position $u(t)$ and the cursor position $y(t)$ on the screen, a controlled element or ``Plant'' intervened, consisting of a single integrator. That is, the cursor moved under rate control. 
The operator's objective was to track a single moving target by commanding cursor motion using their joystick. Both cursor and target motion were visible on the monitor. 
The target moved according to one of three reference signals $r_1(t), \,\, r_2(t)$ and $r_3(t)$ at any given time. Each of the three reference signals was generated as the sum of 10 sinusoids and thus each appeared to move randomly. 
The sum of sinusoids had the form \cite{mcruer1967review,mcruer1965human}
\begin{equation}
    r_i(t) =\sum_{k=1}^{10} A[k] \sin(2 \pi f[k] t)+ \phi_i[k], \quad (i=1,2,3)
\end{equation}
where $f[k]$ were the following multiples of a base frequency of 0.025 Hz: $[6\,\, 10\,\, 15\,\, 23\,\, 37\,\, 57\,\, 97\,\, 154\,\, 289\,\, 527]$, and $A[k]=20, \quad (k=1,...,8), \quad A[k]=1, \quad(k=9,10)$, and $\phi[k]$ were random numbers between 0 and $2\pi$.  

The target on the operator's monitor initially followed one of $r_i(t)$ selected randomly and then switched to another reference $r_j(t)$ if the following four conditions were met: a) $i \neq j$, b) $|r_i(t) - r_j(t)| < \epsilon$, c)$|\dot{r}_i(t) - \dot{r}_j(t)| < \epsilon_v$ and d) $t_s - t < \epsilon_s$, where $\epsilon = 0.05$ and $\epsilon_v = 0.2$ and $\epsilon_s = 1.5625 s$. In condition b), $r_i(t)$ is the position of the reference signal $i$ at time $t$, and ensures the selected target only switches when passing over the next selected target. In condition c), $\dot{r}_i(t)$ is the velocity of the reference signal $i$ at time $t$, which ensures that there is a maximum velocity between the current selected target and the next selected target at the time of switching. Condition d) sets a minimum time between that last target switch, where $t_s$ is the time of the previous switch. 

While the operator saw only one target on their monitor, the supervisor/participant saw all three moving targets on their monitor, moving according to $r_1(t), \,\, r_2(t)$ and $r_3(t)$. Reference signal $r_1$ was shown in green, $r_2$ in yellow, and $r_3$ in pink. The supervisor/participant was tasked with identifying which one of the three reference signals the operator was tracking at any given time during a 90 s trial. The supervisor/participant was instructed to press the corresponding key on the number pad of an alphanumeric keyboard, and a post-it note with the text ``1--green, 2--yellow, 3--pink'' was posted on the supervisor/participant's monitor. While all three references signals on the supervisor/participant's monitor are different colors, the target for the operator to follow remained white regardless of which reference signal the target was following. 

\subsection{Experiment Design}


Each participant was asked to play the role of supervisor under three conditions: \emph{uOff}, \emph{uHaptic}, and \emph{uVisual}. In all conditions three moving targets and one cursor were visible on their monitor. 
In the \emph{uOff} condition, the supervisor/participant's joystick remained stationary as it was not attached to the operator's joystick.  The supervisor/participant 
was asked to judge which of the three moving targets was being tracked by the operator based only on visual information.  They indicated their judgement (as fast as possible) by pressing one of three keys with their left hand.  In the \emph{uHaptic} condition, the supervisor/participant rested their hand on the right joystick which was attached through the bar to the left joystick.  In this case  the operator's command was displayed (through haptic display) to the supervisor/participant's hand. The supervisor/participant was instructed to remain passive and let their hand be driven back and forth by the experimenter (the participant was simply a supervisor and not in control). Supervisors/Participants were asked to place their dominant hand on the joystick in the same posture that they chose while using the operator's joystick during training (see training protocol below). 
The supervisor/participant again indicated their best estimate of which target was being tracked. Lastly, in the \emph{uVisual} condition, the supervisor/participant could see the joystick moving as it was being controlled by the operator (the bar attaching the operator and supervisor/participant joystick was in place), but did not place their hand on the joystick.

\subsection{Experimental Procedure}

Each supervisor/participant received three training sessions before performing under the three experimental conditions. In the first training session the supervisor/participant practiced indicating through a keypress which target they thought was being tracked under the \emph{uOff} condition.  This allowed each supervisor/participant to learn the association between key and the target. 
During the second training session, each supervisor/participant took a seat at the operator station and practiced tracking the moving target with their hand on the operator's joystick.  This training was intended to help each supervisor/participant build a mental model of the plant under control (the integrator).  The third training session consisted of the participant practicing the 
supervisory task in the \emph{uHaptic} condition. 
They placed their hand passively on the joystick at the supervisor/participant's station, which was rigidly attached to the joystick that the operator was using to control the cursor. 

After training the participant took the role of supervisor according to the three conditions defined above. The order of conditions, {\emph{uOff}, \emph{uVisual}, \emph{haptic}}, were randomized for each supervisor/participant to avoid ordering bias. While the experiment was running, the supervisors/participants wore headphones playing white noise to minimize auditory cues from the operator's joystick.

\subsection{Performance Metrics and Statistical Treatments}
The performance of the supervisors/participants was evaluated based on both the percentage of time that the supervisor/participant's input was the same as the target shown to the operator (termed \emph{Target Selection Accuracy} and the delay between the participant's selection and the target shown to the operator (termed \emph{Target Selection Delay}). 
A one-way repeated measures ANOVA was conducted to examine the effect of condition (\emph{uOff}, \emph{uVisual}, \emph{uHaptic}) on \emph{Target Selection Accuracy} and \emph{Target Selection Delay} in 10 participants. Post hoc tests were conducted using the Bonferroni correction.

\section{Results}
Results indicate a benefit to providing a copy of the operator command signal to the supervisor's ability to identify which of the three targets was being tracked by the operator. 

\begin{figure}[h]
\caption{\textbf{Operator Performance:} The target visible to the operator followed reference signal $r_1$ until $t=4.04s$, then $r_2$ until $t=8.84s$, then $r_3$ until $t=10.50s$, then $r_1$ until $t=12.68s$, and thereafter $r_3$. The operator's performance in tracking the selected reference signal can be seen in the dashed black line $y(t)$. The supervisor's keypresses are indicated in the overlay staircase plot. 
}
\centering
\includegraphics[width=1\textwidth]{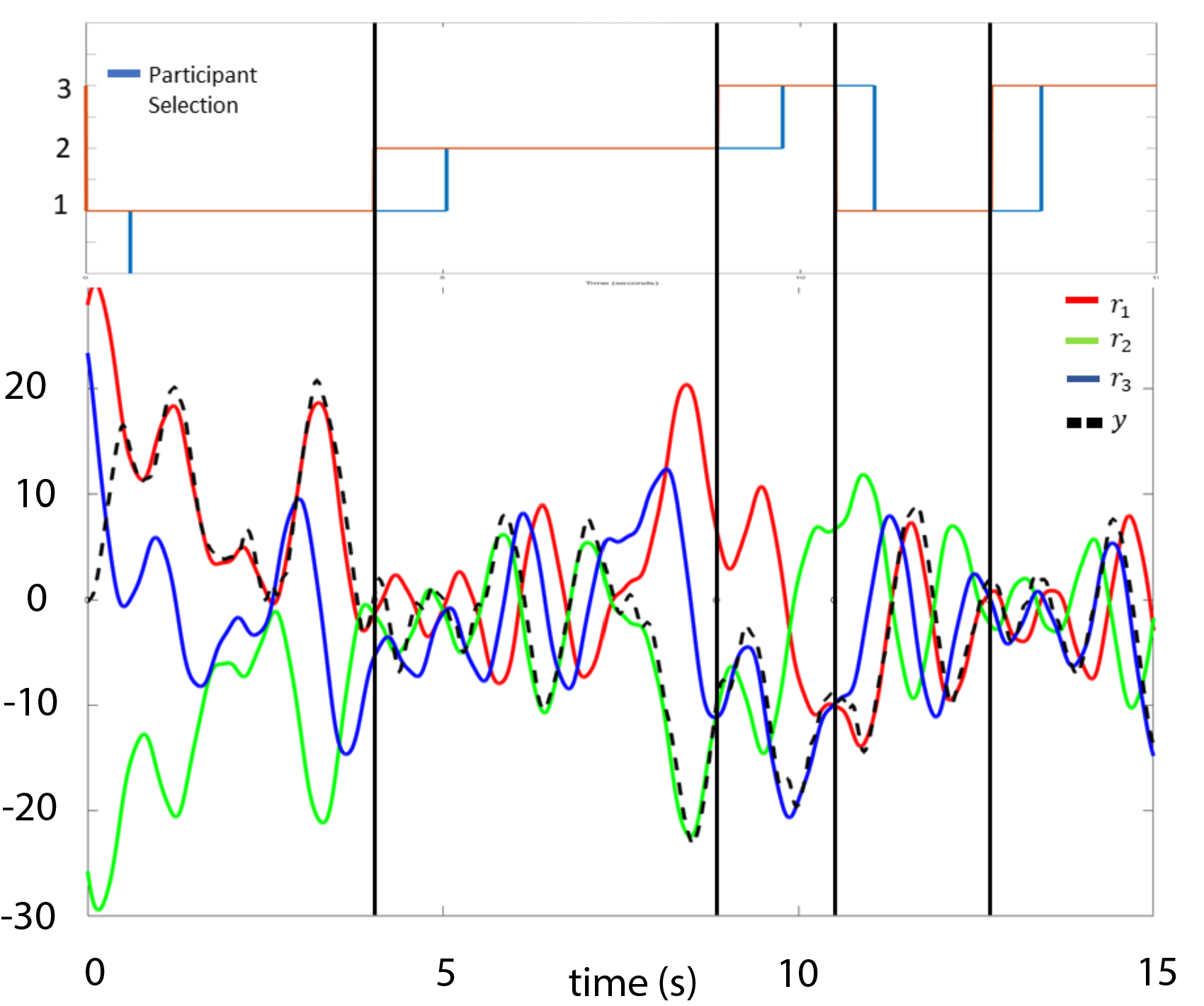}
\label{fig:OperatorTracking}
\end{figure}

Trajectories of all three reference signals and the cursor trajectory produced by the operator are shown in overlay for a 12 s segment of a trial in the \emph{uOff} condition in Fig. \ref{fig:OperatorTracking}.  A staircase plot (sequence of step functions) indicates the timing of the supervisor's keypresses. The tracking performance of the operator and the delay of the supervisor's identification of which one of the three reference signals the operator was tracking can be deduced from these trajectories. 

\subsection{Target Selection Accuracy}

Target Selection Accuracy reflects the cumulative time during which a supervisor correctly identified the target being tracked by the operator as a percentage of the total trial time (90 s). Target Selection Accuracy is shown for the three conditions in a Box and Whiskers plot in Fig \ref{fig:part_perf}.  
On average, as shown in Table \ref{tab_meanstd}, participants achieved higher Target Selection Accuracy in the \emph{uHaptic} condition and lower Target Selection Accuracy in the \emph{uVisual} compared to the \emph{uOff} condition. 

The Target Selection Accuracy was statistically significantly different across conditions, $F(2, 18) = 3.56$, $p < 0.05$, generalized $\eta^2 = 0.046$.
Post-hoc analyses with a Bonferroni adjustment revealed that all the pairwise differences between the \emph{uVisual} and \emph{uHaptic} conditions were statistically significantly different ($p < 0.05$).

\subsection{Target Selection Delay}
\begin{figure}[h]
\caption{Target Selection Accuracy and Target Selection Delay by condition in Box and Whiskers Plots. Significant differences were found in Target Selection Accuracy between the \emph{uVisual} and \emph{uHaptic} conditions. }
\centering
\includegraphics[width=1\textwidth]{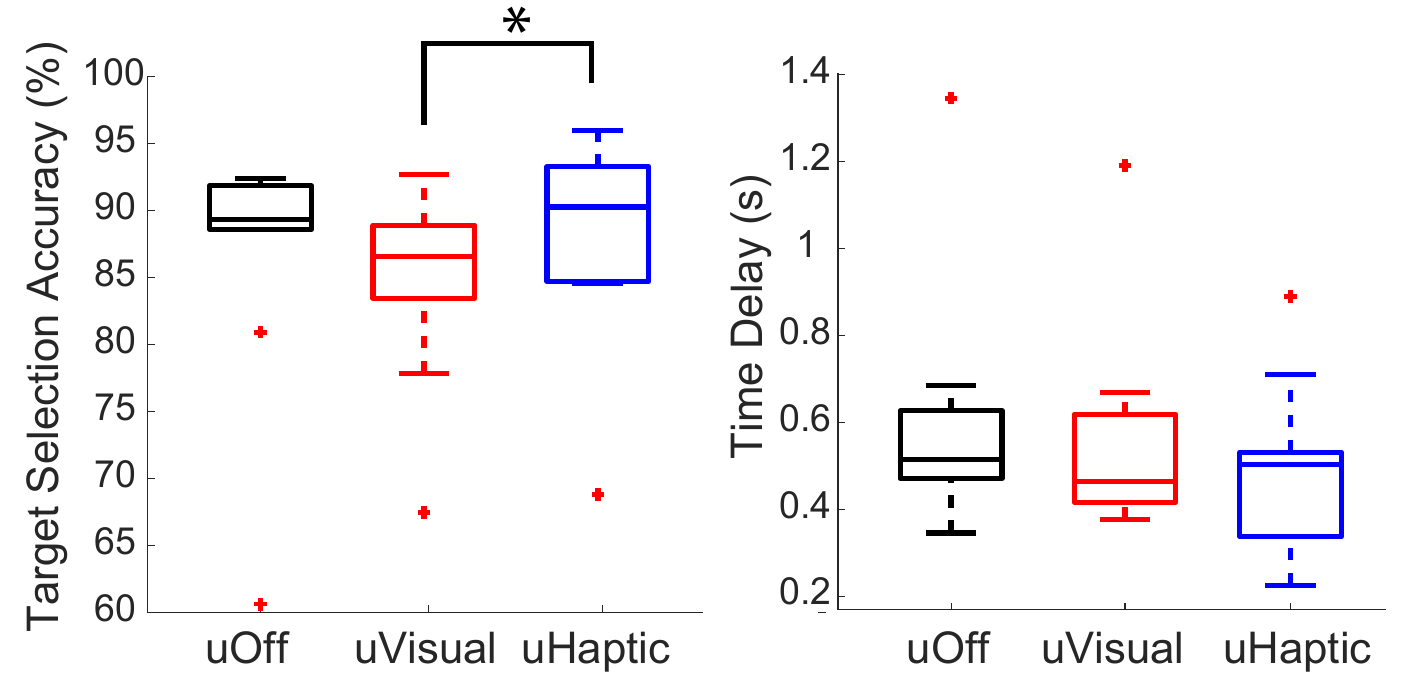}
\label{fig:part_perf}
\end{figure}

The time delay, in seconds, between the occurrence of a switch in the reference signal shown to the operator and a keypress by the supervisor is defined as Target Selection Delay. We computed the Target Selection Delay using the \lstinline{finddelay} function in MATLAB. This function uses the \lstinline{xcorr} function to determine the cross-correlation between each pair of signals. Then the delay estimate is given by the negative of the lag for which the normalized cross-correlation has the largest absolute value.
Target Selection Delay is shown for the three conditions in a Box and Whiskers plot in Fig \ref{fig:part_perf}. 
The participants took less time to select the correct target being tracked in both the \emph{uHaptic} and \emph{uVisual} conditions compared to the \emph{uOff} condition.


\subsection{Operator Error Correlation}
As the performance of the operator significantly affects the performance of the participant, with a correlation coefficient of $0.87$, the average value for each condition of the operator's RMSE error is recorded in Table \ref{tab_meanstd}.

\subsection{Participant Survey}
According to a brief survey conducted at the end of each experiment, participants reported that the haptic feedback made them more confident in their choices. In the \emph{uVisual} condition, participants commented that they either chose to ignore the visual control command (the motion of the joystick in front of them) or that it distracted them from their task. 



\begin{table}[]
    \centering
    \begin{tabular*}{3.1in}{@{\extracolsep{\fill}}||p{0.8in}||p{0.5in}|p{0.5in}|p{0.5in}||} 
         \hline
         $\mu \pm \sigma$ & \textbf{uOff} & \textbf{uVisual} & \textbf{uHaptic}\\ 
         \hline\hline
         Target Selection Accuracy (\%)& $0.8658 \pm 0.0972$ & $0.8469 \pm 0.0739$ & $0.8830 \pm 0.0785$ \\ \hline
         Target Selection Delay (s)& $0.6058 \pm 0.2778$ & $0.5631 \pm 0.2422$ & $0.4938 \pm 0.1955$\\ \hline
         Operator Error Correlation & $1.1452 \pm 0.0956$ & $1.5655 \pm 0.2000$ & $1.4503 \pm 0.1135$\\
         \hline
    \end{tabular*}
    \caption{Mean and Standard Deviation of Performance Metrics by Condition}
    \label{tab_meanstd}
\end{table}

\section{Discussion}
Our results indicate that a supervisor's 
ability to correctly identify the target being tracked and the speed with which they can identify the target being tracked was improved when that supervisor had haptic access rather than only visual access to the control command issued by an operator. According to our results, the supervisor's ability but not speed diminished when they had visual access to the control command relative to having no access to the control command. The lower performance could be attributed partially to higher operator error, as well as the two visual mediums competing for the participants' attention. 


 The delay in the participants' choices decreased in both the \emph{uVisual} and \emph{uHaptic} conditions compared to \emph{uOff}, which would seem to indicate that the participants were more confident in their choice of target at a given time. This result is especially interesting in the \emph{uVisual} condition where the time delay was smaller despite the performance also decreasing. 
 Even though the \emph{uHaptic} had a larger operator error compared to \emph{uOff} (although a slightly larger operator error compared to \emph{uVisual}), the participant percentage performance for \emph{uHaptic} was still the highest of the three conditions. Higher results on the \emph{uHaptic} condition could have been due to the participants' own experience acting as operator in the training sessions, as well as the awareness and training they received to use haptics as a modality to inform their decisions. The training the participants received was intended to lower cognitive load when interpreting haptic feedback, as they could refer to their own experience as controller, which could help explain the short time delays in this condition. 
 
 Larger error on the part of the operator in the \emph{uHaptic} condition can be partially attributed to some amount of resistance (mechanical load) resulting from the participant's hand resting on the opposite joystick. Also, in both the \emph{uVisual} and \emph{uHaptic} cases, there was an added mass to compensate compared to the \emph{uOff} condition when the bar linkage was disconnected.


\section{Conclusion and Future Work}
Planned experiments in which the operator is replaced with an autonomous agent and motor are expected to yield even stronger results. The apparatus shown will include a motor attached to the linear slide, automating the tracking task.  Like the wizard, 
the automation would use feedback to minimize error between the cursor position and a (switching) selected target. By eliminating the human operator, the supervisor no longer needs to accommodate for delays or discrepancies produced by a human wizard. Given the results from the experiments in this paper, future experiments using automation are expected to yield stronger effects among the three conditions. 

Extending this baseline theory to applications, further experiments will include interpretation of first-person footage from an automated ground or automated aerial vehicle. 
We will build on these experiments with contexts that include supervisory control of both air and ground vehicles and contexts in which multiple autonomous vehicles and agents operate simultaneously.

\addtolength{\textheight}{-12cm}   




\section*{ACKNOWLEDGMENT}
We would like to acknowledge Kaleeb Naveed for help on video developemnt and editing and Hannah Baez for help with statistical analysis.


\bibliography{references}

\bibliographystyle{IEEEtran}

\end{document}